\documentclass[%
    aps,
    prd,
    10pt,
    nofootinbib,
    noeprint,
    amsmath,
    amssymb,
    superscriptaddress,
    preprintnumbers,
]{revtex4-2}

\usepackage{graphicx}
\usepackage[%
    colorlinks=true,
    allcolors=blue
]{hyperref}

\usepackage[%
    print-unity-mantissa=false,
    range-phrase=-
]{siunitx}

\usepackage{multirow}

\begin{document}

\title{The tidal response of a relativistic star}

\author{N. Andersson}
\affiliation{Mathematical Sciences and STAG Research Centre, University of Southampton, Southampton SO17 1BJ, United Kingdom}

\author{A. R. Counsell}
\affiliation{Mathematical Sciences and STAG Research Centre, University of Southampton, Southampton SO17 1BJ, United Kingdom}

\author{F. Gittins}
\affiliation{Institute for Gravitational and Subatomic Physics (GRASP), Utrecht University, Princetonplein 1, 3584 CC Utrecht, Netherlands}
\affiliation{Nikhef, Science Park 105, 1098 XG Amsterdam, Netherlands}

\author{S. Ghosh}
\affiliation{Mathematical Sciences and STAG Research Centre, University of Southampton, Southampton SO17 1BJ, United Kingdom}

\date{\today}

\preprint{INT-PUB-25-027}

\begin{abstract}
We develop a fully relativistic approach for determining the response of a compact star to a time/frequency dependent (tidal) environment. The strategy involves matching the solution for the linearised fluid dynamics in the star's interior to the spacetime perturbations in the near-zone surrounding the body, along with an identification of the tidal driving and the star's response. Notably, this identification is exact in Newtonian gravity and we provide strong evidence that it remains robust also in the relativistic case. The argument does not involve a sum over the star's quasinormal modes and hence circumvents one of the obstacles that have held up the development of models for relativistic tides. Numerical results are provided, at the proof-of-principle level, for a realistic matter equation of state from the BSk family, including composition stratification leading to the presence of low-frequency gravity modes. We also sketch the connection with the field-theory inspired approach to the problem, in which the tidal response is expressed in terms of asymptotic scattering amplitudes. We argue that progress in this direction is essential in order to complete the description of tides in relativistic stars.
\end{abstract}

\maketitle

\section{Introduction and scope}

Gravitational-wave astronomy has the potential to constrain the nature of matter under the extreme densities and pressures of neutron star cores. The arguments for this are well exercised in the celebrated case of GW170817 \cite{PhysRevLett.119.161101,2017ApJ...848L..13A} for which the inferred range for the tidal deformability---a representation of the star's static tidal response---allows the inference of the neutron star radius \cite{2018PhRvL.121p1101A,2018PhRvL.121i1102D}. The associated error bars are similar to the ones for the sample of neutron star radii obtained from x-ray pulse profile modelling and NICER data \cite{2024ApJ...971L..19R}. As the sensitivity of the gravitational-wave interferometers is improved, in first instance to the LIGO A+ level and eventually to the next-generation instruments, Cosmic Explorer \cite{evans2023cosmicexplorersubmissionnsf} and the Einstein Telescope \cite{abac2025scienceeinsteintelescope}, the expectation is that the tidal constraints will become more precise \cite{2025PhRvD.112f3020K} and that additional tidal aspects---e.g., the mode resonances associated with the dynamical tide---will come within reach \cite{2016PhRvL.116r1101H,2016PhRvD..94j4028S,2023PhRvD.108f3026K,2024PhRvD.110b4039Y,2025MNRAS.542.1375P}. This development promises to be transformative as dynamical aspects shed light on the both the composition \cite{2018PhRvD..97b3016A,2021MNRAS.506.2985K,2023PhRvD.108d3003H,2025MNRAS.536.1967C,2025PhRvL.135h1402C} and state of matter \cite{2021MNRAS.504.1273P,2022MNRAS.514.1494P} at high densities; information that is difficult to glean in any other way.

A typical neutron star binary system will spend the last 15 minutes or so in the sensitivity band of future ground-based interferometers (above 10 Hz) \cite{evans2023cosmicexplorersubmissionnsf,abac2025scienceeinsteintelescope}. The early phase of the inspiral signal allows for the extraction of accurate masses and spins for each of the two bodies, while the late stages of evolution (above a few 100 Hz) involve the mutually induced tidal interaction. A central promise of next-generation era gravitational-wave astronomy involves using the fine print tidal signature to constrain high-density physics. However, to realize this potential we need to develop models of dynamical tides in neutron-star binaries to the level of precision required by future instruments, providing a robust description of the signal emitted as a system evolves through the detector sensitivity band and sharp statements regarding the observability of fine print features connected with the composition and state of matter. 
Such models inevitably require a fully relativistic description of the response of the stellar matter to the external tidal driving. However, developments in this direction have long been hampered by technical issues (see \cite{2024PhRvD.109f4004P} for a concise summary of relevant points). While formal neutron-star perturbation theory is well developed, the connection to the tidal problem  remains a challenge.  

As a result, work on dynamical neutron star tides was, until fairly recently, exclusively carried out within the framework of Newtonian gravity \cite{1994MNRAS.270..611L,1995MNRAS.275..301K,2020PhRvD.101h3001A,2024MNRAS.527.8409P}. This has the advantage that the dynamical tide can be expressed as a sum over the star's oscillation modes, a representation that is known to be complete as the perturbation problem is Hermitian. No such representation is (formally) expected to exist in the relativistic case. One obvious reason for this is that the star's oscillation modes are “quasi-normal” (including  damping due to gravitational-wave emission) and hence the perturbation problem is not conservative. Currently, progress is being made by exploring a range of approximations, from the post-Newtonian models in \cite{2025CQGra..42m5014G,2025arXiv250406918Y} to the relativistic Cowling approximation \cite{2021MNRAS.506.2985K,2025MNRAS.536.1967C}, low-frequency approximations \cite{2024PhRvD.109f4004P,2024NatAs...8.1277R}, and hybrid models involving an approximate tidal driving \cite{2025arXiv250710693A}.

Acknowledging the technical difficulties of the problem, we set ourselves a modest target. We focus on the \emph{response} of a relativistic star to a time/frequency dependent (tidal) environment, hoping to demonstrate that this problem can be approached in a fairly straightforward manner. This will admittedly leave us with work to do---e.g. in order to link our results to the evolution of the binary orbit---but it still represents useful progress.

\begin{figure}
    \centering
       \includegraphics[height=8cm]{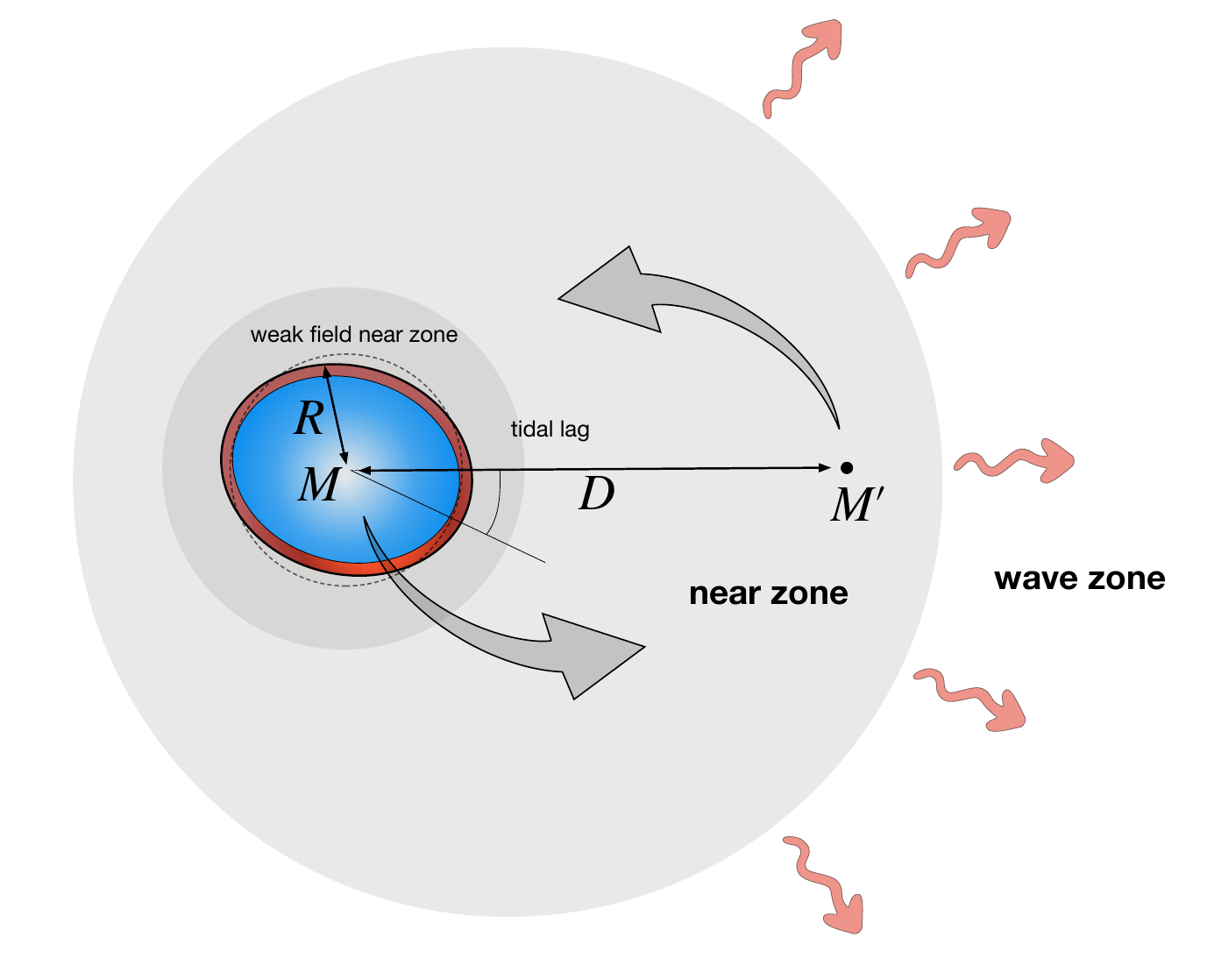}
    \caption{A schematic illustration of the tidal problem. A star with mass $M$ and radius $R$ is tidally deformed by a binary companion with mass $M'$. The orbital separation is $D$. Our focus is on the star's response to the tidal interaction, a problem that is naturally explored in a body centered coordinate system associated with the star \cite{2013PhRvD..88b4046V, 2024NatAs...8.1277R} and the weak-field near zone in the vicinity of the star. In addition, the orbital problem involves the near zone that envelops the two bodies and the emitted gravitational radiation, a problem that involves the wave zone.}
    \label{tidefig}
\end{figure}

In order to understand the context, consider the issue of tides in the binary neutron star problem \cite{2005PhRvD..71d4010R,2013PhRvD..88b4046V,2016PhRvD..94j4028S,2021PhRvD.103f4023P,2024PhRvD.110d4041R}. 
The tidal problem is naturally separated into three stages, see Figure~\ref{tidefig}. First, we need to work out how a given relativistic star responds to the tidal driving (inner problem). This involves placing the star in an external (possibly time varying) tidal environment and quantifying the induced perturbations in the stellar fluid and the spacetime metric. This calculation is naturally carried out in a body centred coordinate system associated with the star \cite{2013PhRvD..88b4046V, 2024NatAs...8.1277R}. The second step involves working out the impact on the orbital evolution (outer problem), assessing to what extent the tidal imprint may be extracted from an observed signal and how the relevant parameters of interest enter the problem (waveform problem). We will only consider the first of these problems here.

Given that (current) numerical simulations are able to track, at best, the last few tens of binary orbits before merger \cite{2025PhRvL.135n1403K} (and noting that current simulations are associated with somewhat uncontrolled systematic errors \cite{2022PhRvD.105f1301K,2025PhRvD.111b3049G}), the low-frequency tidal problem inevitably involves approximations. This, in turn, relies on a separation of scales and (typically) a matching to the tidal environment in a region close to the star, as indicated in Figure~\ref{tidefig}. First, the binary problem can be separated into a near zone and a wave zone. In the latter, gravitational perturbations are represented by waves. Meanwhile, in the near-zone, where the two bodies reside, the gravitational perturbations can be viewed as slowly varying. Second, as long as the binary separation is large enough, the tidal interaction can be adequately described perturbatively. In essence, each star is tidally deformed by the presence of the binary partner, but it is assumed that the interaction is weak enough that it induces a suitably small perturbation (e.g., linear) of star's equilibrium state. This then defines the weak-field near zone (indicated in Figure~\ref{tidefig}). This region is key to the argument we will develop.

The weak-field near zone is also central to approaches to the outer problem and the connection to the binary evolution. It is worth noting that our approach interfaces naturally with both the world-line field theory approach  and the effort to describe the tidal problem in terms of scattering amplitudes \cite{2021PhRvD.104l4061C,2023PhRvL.130i1403I,2024PhRvD.109f4058S,2024PhRvD.110j3001S,2025arXiv250623176M}  and the self-force programme \cite{2023PhRvD.108b4041W,2024PhRvD.110h4023W}. In the former, the tidally perturbed body is represented by a set of induced multipole moments which can be connected to asymptotic scattering amplitudes. In the latter, one would envisage directly matching the tidally deformed body to the external spacetime. Both approaches involve a solution to the problem in the weak-field near zone. This motivates our (fairly pragmatic) approach to the tidal response problem.

\section{Motivation: The Newtonian problem}

To set the scene for the proceedings, let us first consider the tidal problem in Newtonian gravity (following \cite{1994MNRAS.270..611L,1995MNRAS.275..301K,2020PhRvD.101h3001A,2024MNRAS.527.8409P}). Focussing on the gravitational field in the local coordinate system associated with the primary star,
the gravitational potential associated with the binary companion, $\chi$, is simply given by a solution to the homogeneous problem 
\begin{equation}
    \nabla^2 \chi = 0,
\end{equation}
sourced by the matter distribution of the partner and (obviously) encoding its evolving (relative) location.
This potential deforms the matter in the primary star, leading to a perturbed mass density, $\delta \rho$, and (in turn) a perturbed gravitational potential, $\delta \Phi$, determined by 
\begin{equation}
\nabla^2 \delta \Phi = 4\pi G \delta \rho.
\end{equation}
Evidently, because we assume the problem is in the linear regime, we may add the two potentials and solve the single Poisson equation
\begin{equation}
    \nabla^2 U = 4 \pi G \delta \rho,
\end{equation}
with $U=\delta \Phi + \chi$. This solution is valid only in the regime where $\chi$ is small compared to the gravitational potential of the background star, $\Phi$, given by
\begin{equation}
    \nabla^2 \Phi = 4 \pi G \rho, \quad \implies \quad \Phi = - \frac{G M}{r}, \quad \mbox{for } r \geq R,
\end{equation}
with $M$ and $R$  the mass and radius of the equilibrium body, respectively.

We now focus our attention on the vacuum exterior to the tidally deformed star,   far enough away  from the companion for the tidal field to be considered a small perturbation. In essence, we operate in the weak-field near zone indicated in Figure~\ref{tidefig}. In this region, obviously corresponding to $r \geq R$, we simply need to solve the homogeneous version of Laplace's equation. After expanding in spherical harmonics, we have the general solution
\begin{equation}
    U(t, r, \theta, \varphi) = \sum_{l, m} U_l(t, r) Y_l^m(\theta, \varphi), 
\end{equation}
with
\begin{equation}
U_l = \frac{A_l}{r^{l + 1}} + B_l r^l.
\end{equation}
It is easy to see that the combined potential $U_l$ separates neatly into a decaying component, naturally identified with the perturbed gravitational potential of the star;
\begin{equation}
    \delta \Phi_l = \frac{A_l}{r^{l + 1}},
\end{equation}
and a growing component, which must arise from the tidal potential;
\begin{equation}
    \chi_l = B_l r^l.
\end{equation}
There are no surprises here. Given that the Newtonian field equation is linear, the two terms can be decoupled.

Moreover, it is straightforward to connect our solutions to (a generalised version of) the tidal Love number, $k_l$, defined at the surface of the star, $r = R$ (see, for example \cite{2024MNRAS.527.8409P});
\begin{equation}
    \delta \Phi_l(t, R) = 2 k_l(t) \chi_l(t, R).
    \label{kldef}
\end{equation}
That is, we may identify
\begin{equation}
    k_l = {A_l \over 2 R^{2l+1} B_l},
\end{equation}
and express the exterior solution as 
\begin{equation}
    U = \sum_{l, m} \left[1 +  2 k_l \left( \frac{R}{r} \right)^{2 l + 1}  \right] \chi_l Y_l^m.
\end{equation}
This relation neatly highlights that we can make progress on determining the Love number (which is a relative measure) without knowing the overall amplitude (dictated by the strength of the tidal potential, $\chi_l$). In essence, we can quantify key aspects of the star's tidal response without establishing the connection to the binary orbit and its evolution. It will be useful to keep this in mind later. 

For completeness, it is worth recalling that the main observable associated with the neutron star tide is not the Love number, but rather the (also dimensionless) tidal deformability. The dominant (quadrupole) contribution to the gravitational-wave phasing is expressed in terms of
\begin{equation}
\Lambda_2  = {2\over 3} \left( {R c^2 \over GM} \right)^5 k_2 .
\end{equation}

Moving on, it should be clear that the expressions we have provided so far apply equally to a dynamical context. We did not need to restrict the time dependence of the tidal field. If the problem is time evolving, as it generally will be, then it is natural to work in the frequency domain (making use of an integral transform, associated with a frequency $\omega$, say). For slowly evolving, circular motion it then follows immediately that the driving frequency is $\omega = m\Omega$, with $\Omega$ the Keplerian orbital frequency. Denoting the frequency domain variables by hats and noting that the matching of the gravitational potential and its derivative is carried out at the star's surface we have
\begin{gather}
    l \hat{U}_l - r \partial_r \hat{U}_l = (2 l + 1) \delta \hat{\Phi}_l, \\
    (l + 1) \hat{U}_l + r \partial_r \hat{U}_l = (2 l + 1) \hat{\chi}_l. \label{eq:Relation}
\end{gather}
From these relations, we readily arrive at 
\begin{equation}
        k_l(\omega) = \frac{1}{2} \left[ l\hat{U}_l(\omega, R)  -  R \partial_r \hat{U}_l(\omega, R) \right] \left[ (l+1) \hat{U}_l(\omega, R) + R \partial_r \hat{U}_l(\omega, R) \right]^{-1}.
    \label{eq:Love-alt}
\end{equation}
This expression provides the effective, frequency dependent, Love number. It is worth emphasizing that the argument does not involve explicitly separating the two potentials in the stellar interior. It is also worth noting that, as we later consider how the result varies with the frequency, we treat both the frequency $\omega$ and the stellar radius $R$ as independent variables required to determine $\hat U_l (\omega,R)$.

The argument we have outlined suggests the following strategy, notably not relying on the fluid perturbation equations being Hermitian or involving an explicit mode sum. The first step involves solving the usual stellar perturbation problem for a general (real) frequency $\omega$. Noting that  $\hat U_l$  enters the problem in the same way as $\delta \hat \Phi_l$ this leads to a solution $(\hat \xi_l^i, \hat U_l)$ (where $\hat \xi_l$ is the  displacement vector that represents the fluid motion in Lagrangian perturbation theory \cite{1978ApJ...221..937F}). This solution is required to satisfy the condition of regularity at the star's centre and be such that the Lagrangian perturbation of the pressure vanishes at the surface. Given this solution we have access to $\hat U_l(\omega,R)$ and its derivative, and hence we can evaluate $k_l$ for the assumed frequency. The frequency dependence of the tidal response follows by repeating the procedure for a desired range of frequencies (with the required numerical resolution). This is straightforward.

At this point we should make two remarks. First, as a sanity check, it is worth keeping in mind that the expression in \eqref{eq:Love-alt} must limit to the usual Love number in the static ($\omega\to0$) limit. Second, in the absence of a tidal field  we have $\hat{\chi}_l = 0$, and Eq.~\eqref{eq:Relation} then provides the  boundary condition that ensures that the gravitational potential decays away from the star. The corresponding solutions represent the free oscillations of the star for a discrete set of frequencies $\omega_n$ (with $n$ labelling each mode). This connects the tidal response problem with asteroseismology \cite{1998MNRAS.299.1059A}.

It is easy to see that \eqref{eq:Love-alt} will be singular for $\omega=\omega_n$. This is, indeed, what happens also for the tidal mode sum. In fact, it is easy to connect the matching argument to the mode-sum result. Close to a given mode frequency $\omega_n$ (but far enough away from all other modes) we can Taylor expand the denominator in \eqref{eq:Love-alt}. This leads to
 \begin{multline}
   (l+1)\hat{U}_l(\omega, R) + R \partial_r \hat{U}_l(\omega, R) \approx \underbrace{(l+1)\hat{U}_l(\omega_n, R) + R \partial_r \hat{U}_l(\omega_n, R)}_{=0} \\
   + (\omega - \omega_n)  \left\{(l+1) \partial_\omega \hat{U}_l(\omega, R)  + R   \partial_r[\partial_\omega \hat{U}_l(\omega, R)] \right\}_{\omega=\omega_n}.
 \end{multline}
 We may compare this result to the anticipated mode-sum expression---assuming that the normalisation of the mode solutions is taken to be $MR^2$ and rescaling the overlap internal in such a way that $Q_n = M R^l \tilde Q_n$, see for example, \cite{2025MNRAS.542.1375P};
\begin{equation}
    k_{l} =  \frac{2 \pi }{2 l + 1 }
            \sum_{n} {\tilde Q_{n}^2  \over 
            \tilde \omega_{n}^2 - \tilde \omega^2}.
            \label{modesum1}
\end{equation}
We have also introduced the usual dimensionless frequency
\begin{equation}
    \omega^2 = \tilde \omega^2 {G M \over R^3}.
\end{equation}

It follows that
\begin{equation}
\tilde Q^2_n =  -  {2l+1 \over \pi} \tilde \omega_n \lim_{\tilde \omega \to \tilde \omega_n} ( \tilde \omega -\tilde \omega_n) k_l(\tilde\omega) =  - {2l+1 \over \pi} \tilde \omega_n \mathrm{Res}\, k_l(\tilde \omega_n) .
\end{equation}
The expression for the overlap integral obtained from matching the potentials at the star's surface then takes the form
\begin{equation}
\tilde Q^2_n =  -  {2l+1 \over 2\pi} \tilde \omega_n \left[ l\hat{U}_l(\tilde\omega_n, R)  -  R \partial_r \hat{U}_l(\tilde \omega_n, R) \right] \left\{ (l+1) \partial_\omega\hat{U}_l(\tilde \omega, R)  + R   \partial_r [\partial_\omega\hat{U}_l(\tilde \omega, R)] \right\}_{\tilde \omega=\tilde \omega_n}^{-1} .
\label{matchingQ}
\end{equation}

\section{Implementation in Newtonian gravity}

Let us put the proposed strategy to the test by implementing it, in the first instance, for the case of an incompressible star. For this problem, the solution to the perturbation equations is straightforward (see, for example, Chapter~13 in \cite{2019gwa..book.....A}), and we find that the solution for the external potential takes the form
\begin{equation}
    \hat U_l(\omega, r) =  \left\{ {1\over l R^{l-1}} \left[ \omega^2 - {GM  \over R^3} {2l \left( l - 1 \right) \over 2l+1}\right]    r^l - {4\pi G \rho R \over 2l+1}  \left( {R \over r} \right)^{l+1} \right\} \hat \xi_l (R) \ .
    \label{Ulext}
\end{equation}
where $\hat \xi_l(R)$ represents the radial component of the displacement vector at the star's surface $(r=R)$ and $\rho$ is the constant (mass-)density. In the absence of a tidal interaction (with $\hat \chi_l=0$), the exterior potential must decay as $r\to \infty$. We see that this requires (as long as $\hat \xi_l(R)\neq 0$)
\begin{equation}
   \omega^2 =  \omega_l^2 =  {2l \left( l - 1 \right) \over 2l+1} {GM  \over R^3} \ .
\end{equation}
These frequencies represent the free (f-mode) oscillations 
of the incompressible star. That is, 
we have
\begin{equation}
    \tilde \omega_l^2 = {2l \left( l - 1 \right) \over 2l+1}.
\end{equation}
It then follows that we should identify
\begin{equation}
   \hat \chi_l =  {1\over l R^{l-1}} \left( \tilde \omega^2 - \tilde\omega^2_l \right)   {G M \over R^3} \hat \xi_l (R)   \ ,
\end{equation}
directly relating the shape of the star's surface to the external tidal potential.
Moreover, using our solution in \eqref{eq:Love-alt} we find that 
\begin{equation}
    k_l =   {\tilde \omega_l^2 - l \over  2(\tilde \omega^2 - \tilde \omega_l^2)}.
\label{kleff1}
\end{equation}
This agrees with the result obtained in \cite{2020PhRvD.101h3001A}.  We also see that, in the limit $\tilde \omega\ll \tilde \omega_l$ we have
\begin{equation}
k^\mathrm{eq}_l = {3\over 4 (l-1)} \ ,
\end{equation}
which is the expected result for the static tide of an incompressible body \cite{2014grav.book.....P}.

The fact that the proposed matching scheme reproduces the known results for a simple toy model is reassuring (although not in any way surprising). The most useful intuition we gain from the exercise relates to the singularity in the tidal response associated with the star's oscillation modes. Our calculation shows explicitly how the singularity arises due to the fact that an oscillation mode---the natural response of the star---may be excited without the need for an external driving agent. It is useful to keep this in mind for later.

However, the incompressible model does not shed particular light on the behaviour for more realistic matter models; in the first instance, when the fluid is compressible and composition stratification leads to a family of low-frequency gravity modes \cite{1992ApJ...395..240R}. Fortunately, the generalisation to this case is straightforward.
Formally, the only things that change are: (i) the interior solution must now account for internal density perturbations ($\delta \hat \rho \neq 0$),  and (ii) one would typically have $\rho(R)=0$, so the surface boundary condition on the pressure perturbation simplifies to $\delta \hat p=0$ at $r=R$. Neither of these changes impact on \eqref{eq:Love-alt}. The only adjustment we need to make is that the perturbation equations have to be solved numerically. 

\begin{figure}
\centering\includegraphics[width=0.6\textwidth]{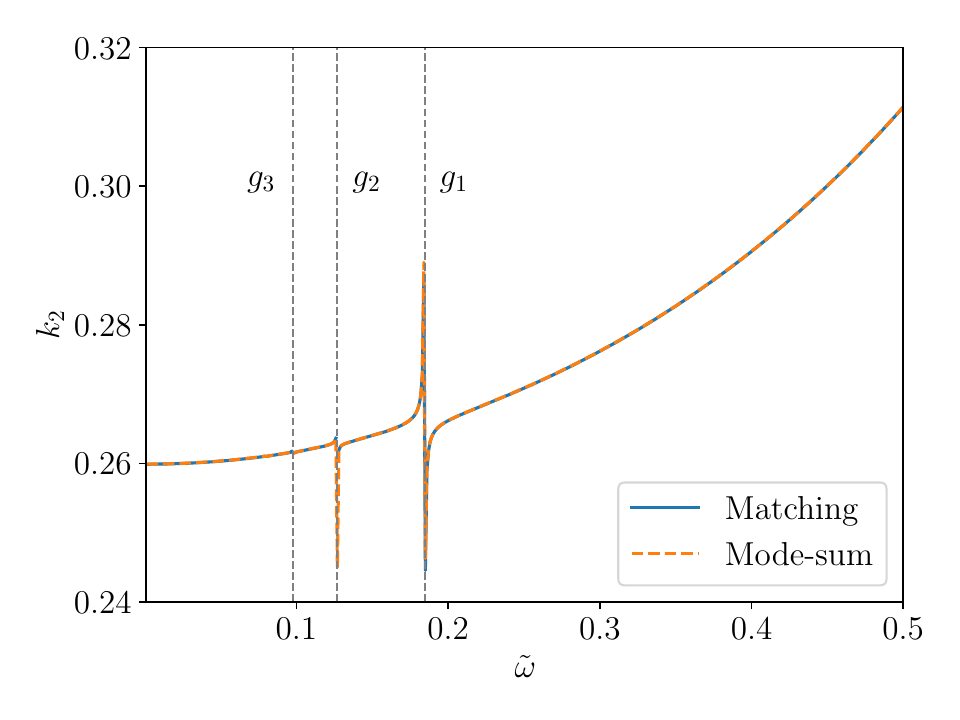}
    \caption{A comparison of the effective Love number obtained from (i) the standard mode sum  \eqref{modesum1} (orange+dashed) and (ii) the proposed matching approach \eqref{eq:Love-alt} (blue+solid). The results, which correspond to a compressible model with $\Gamma=2$ and the $\Gamma_1=2.05$ case from \cite{2021MNRAS.504.1273P}, are in perfect agreement. The  singularities associated with the first two gravity modes are particularly prominent in the tidal response. It is also worth noting that the effective Love number approaches the static result, $k_2\approx 0.2599$, in the low-frequency limit. }
    \label{fig:DeltaPhiTest}
\end{figure}

As an illustration, we rework one of the polytropic models from \cite{2021MNRAS.504.1273P}. Specifically, we assume the equilibrium background model is obtained from 
\begin{equation}
    p = K\rho^\Gamma,
\end{equation}
with $\Gamma=2$, while the Lagrangian fluid perturbations are such that
\begin{equation}
    \Delta p = {p\Gamma_1 \over \rho} \Delta \rho .
\end{equation}
If $\Gamma_1\neq \Gamma$ this model supports gravity g-modes. Drawing on the results for the corresponding (dimensionless) mode frequencies ($\tilde \omega_n$) and overlap integrals ($\tilde Q_n$), reproduced in Table~\ref{OverlapTable}, we can compare the result we obtain from \eqref{eq:Love-alt} to the standard Newtonian mode sum \eqref{modesum1}.
Typical results are provided in Figure~\ref{fig:DeltaPhiTest}. Evidently, the two approaches to the tidal response problem are  in perfect agreement. It is also notable that the effective Love number approaches the expected result, $k_2\approx 0.2599$, in the static limit. This demonstration provides further confidence in the strategy we are promoting.

Before we move on, it is worth considering the actual form of the tidal response we infer from the matching argument. 
While the two representations---the matching expression from \eqref{eq:Love-alt} and the mode sum from \eqref{modesum1}---agree perfectly, they are  functionally  different. We know that a general neutron star model formally supports an infinite set of mode solutions for each value of $l$. Moreover, for non-rotating stars we know that these modes are degenerate in $m$ and the mode frequencies appear in symmetric pairs $\pm \omega_n$. The latter follows immediately from the fact that the (non-dissipative) perturbation equations only involve $\omega^2$. What does this mean for our calculation? The amplitude $B_l$  vanishes, and hence $k_l(\omega)$ must be singular, at each mode frequency. The final result should therefore take the form of a product rather than a sum. Of course, we can always recast the result as a sum by means of partial fractions. However, in practice, there is no need to carry out this exercise. The connection we need is provided by the argument that led to \eqref{matchingQ}.
This suggests an additional avenue for testing the numerical implementation. We can compare  the overlap integrals obtained in the usual way to the results from the matching argument. In the Newtonian problem, the two descriptions should be one-to-one, given that the modes are expected to form a complete basis. Given this, the good agreement between the two calculations brought out by the results in Table~\ref{OverlapTable} does not come as a surprise.

\renewcommand{\arraystretch}{1.3}
\begin{table}[ht]
\normalsize
\centering
\caption{Quadrupole ($l=2$) mode frequencies and overlap integrals for $\Gamma=2$ polytropic models and (constant) stratification given by two different values for $\Gamma_1$. The usual calculation, see \cite{2021MNRAS.504.1273P}, leads to the provided values for $\tilde Q_n$ while the matching argument and \eqref{matchingQ} provides a direct comparison. As anticipated, given the completeness of the Newtonian mode sum, the numerical results are in very good agreement.}
\begin{tabular}{|c|c|c|c|c|c|c|}
\hline
\multicolumn{1}{|c}{\multirow{2}{*}{Mode}} & \multicolumn{3}{|c|}{$\Gamma_1=2.05$}& \multicolumn{3}{c|}{$\Gamma_1=2.1$}\\
\cline{2-7}
\multicolumn{1}{|c|}{} & $\tilde\omega_n$ & $\tilde Q_n$ & Matching & $\tilde \omega_n$ & $\tilde Q_n$ & Matching\\
\hline
\multirow{6}{*}{}  
                                $f$ & 1.2277 &	5.58E-01 &	5.38E-01 &1.2277	&5.58E-01 &	5.56E-01\\
                                $g_1$ &0.1845 &	1.77E-03&	1.77E-03 &0.2566	&3.50E-03&	3.50E-03 \\
                                $g_2$ & 0.1270	&4.24E-04	&4.23E-04&0.1770	&8.36E-04	&8.34E-04\\
                               $g_3$ & 0.0975	&1.25E-04&	1.25E-04&0.1361&	2.46E-04	&2.48E-04 \\
                               $g_4$ &  0.0794&	4.14E-05&	4.09E-05&0.1109	&8.13E-05&	8.14E-05 \\
                               
            \hline
\end{tabular}
\label{OverlapTable}
\end{table}

\section{Proof of principle: The relativistic case}

Having established the near-zone matching strategy for the tidal response we  want to explore to what extent we can implement the idea in general relativity. The key point is that Eq.~\eqref{eq:Love-alt} demonstrates that we are able to calculate $k_l$ without the need of a mode sum. This is crucial  because is suggests a way of circumventing the fact that the relativistic mode problem is not Hermitian, an issue which has hampered progress in describing dynamical tides of relativistic stars \cite{2024PhRvD.109f4004P}. 

Two issues immediately manifest in general relativity. First, the Einstein field equations are non-linear. This means that we need to be careful in disentangling the ``tidal potential'' from the ``response'' induced in the stellar fluid. However, this problem does not seem critical as long as we focus on the regime where the tidal interaction is weak. Assuming that the tidal problem is linear, the tidal driving may be considered a perturbation of the star (just as in the Newtonian case) and there is, in fact, no need to separate the tidal driving from the star's response inside the star. The perturbation problem simply involves solving the linearised Einstein field equations (assuming geometric units, $G = c = 1$)
\begin{equation}
    \delta G_{a b} = 8 \pi \delta T_{a b},
\end{equation}
where the spacetime curvature associated with the metric $g_{ab}$ is encoded in the Einstein tensor $G_{ab}$ and the matter is represented by the stress-energy tensor $T_{ab}$. As in the Newtonian calculation, Eulerian perturbations are represented by $\delta$. 
Here is the key point: As long as the tidal problem is in the linear regime---essentially, when the induced metric perturbations $h_{ab}$ are small (in a suitable sense) compared to the gravitational field of the background---the problem we need to solve remains the same as for a perturbed isolated star. The only difference is that the interior solution must match to an exterior that accounts for the tidal environment. The fact that the difference between the two problems is encoded in the boundary condition should immediately remind us of the  matching argument from the Newtonian problem. In essence, the perturbed metric, $h_{a b}$, is analogous to the combined potentials, $U = \delta \Phi + \chi$, in the Newtonian case and we already know that we can solve the tidal problem without separating these two contributions inside the star. There is no obvious reason why the same strategy should not work in relativity.

The second issue presents more of a challenge.
In the vacuum outside the star, the problem obviously reduces to
\begin{equation}
    \delta G_{a b} = 0.
\end{equation}
It is well-known that this problem has two degrees of freedom, representing gravitational waves. The corresponding problem is familiar from black-hole perturbation theory and commonly represented by either the Regge-Wheeler equation or the Zerilli equation \cite{2019gwa..book.....A}. These equations are known to be gauge invariant. Far away from a perturbed star, the two solutions to the linearised Einstein equations represent out- and ingoing gravitational waves. 
 An isolated star can obviously only source outgoing waves, a boundary condition that corresponds to the star's oscillation modes  being damped by gravitational-wave emission and hence quasi-normal in the established parlance. The mode frequencies need to be complex in order to reflect the gravitational-wave damping and the perturbation problem cannot be Hermitian. Strictly speaking, one would not expect a mode-sum representation for the relativistic star to exist (or at least not be complete). We will need to make approximations. As far as we are aware, the best efforts to date involve either (i) developing a post-Newtonian description of the tidal response  \cite{2025CQGra..42m5014G,2025arXiv250406918Y}, (ii) working in the relativistic Cowling approximation \cite{2021MNRAS.506.2985K,2025MNRAS.536.1967C} or (iii) matching the stellar perturbations to a post-Newtonian tidal environment \cite{2024PhRvD.109f4004P,2024NatAs...8.1277R}. None of these strategies is fully relativistic.

In general relativity (and Schwarzschild coordinates), the metric potential is
 \begin{equation}
        g_{t t} = - e^\nu =  - 1 + \frac{2 M}{r},
 \end{equation}
 where $M$ represents the gravitational mass insider radius $r$, 
effectively encodes the gravitational potential. This follows immediately from the Newtonian limit, where (explicitly including the speed of light $c$ to highlight how the Newtonian limit emerges)
\begin{equation}
    g_{t t} = - \left( 1 + \frac{1}{c^2} 2 \Phi \right).
\end{equation}
Given this correspondence with the Newtonian case, it is natural to focus a discussion of the tidal response on the corresponding metric perturbation
\begin{equation}
h_{t t} = - e^\nu H_0 Y_l^m e^{i \omega t},
\end{equation}
where we again assume that we work in the frequency domain.
In Regge-Wheeler gauge the equation that governs $H_0$ was written down some time ago, see for example \cite{1969ApJ...158..997T,1992PhRvD..46.4289K,1997PhRvD..56.2118L}. In vacuum---which is all we  need for the present discussion---the equations provided in \cite{1997PhRvD..56.2118L} (for a non-rotating star) take the form
\begin{equation}
    r^2 \frac{d^2 H_0}{dr^2} + (2 + r \eta_1) r \frac{dH_0}{dr} + [(r \omega)^2 e^{- \nu} - l (l + 1) + r^2 \eta_2] e^\lambda H_0 = 0,
    \label{eq:Exterior}
\end{equation}
with
\begin{align}
    \eta_1 &= \frac{1}{2} \left( \frac{d\nu}{dr} - \frac{d\lambda}{dr} \right) + \frac{2}{r} \left( 2 - r \frac{d\nu}{dr} \right) (\beta_1 - 1), \\
    \eta_2 &= e^{-\lambda} \left[ \frac{d^2 \nu}{dr^2} + \left( \frac{d\nu}{dr} \right)^2 + \frac{1}{2 r} \left( 2 - r \frac{d\nu}{dr} \right) \left( 4 \beta_2 - \frac{d\nu}{dr} + \frac{d\lambda}{dr} \right) \right], \\
    \beta_1 &= \Delta \left\{ - [2 (r \omega)^2 e^{-\nu} - (l - 1) (l + 2)] l (l + 1) e^\lambda + 2 (r \omega)^2 e^{-\nu} \left( 2 - r \frac{d\nu}{dr} \right) \right\}, \\
\begin{split}
    \beta_2 &= \Delta \bigg\{ r \omega^2 e^{-\nu} \left( 2 - r \frac{d\nu}{dr} \right) \left[ l (l + 1) e^\lambda - 2 \left( 1 - r \frac{d\nu}{dr} \right) \right] \\
    &\qquad\quad- [(l - 1) (l + 2) - 2 (r \omega)^2 e^{-\nu}] \left[ - l (l + 1) \frac{d\nu}{dr} + 2 r \omega^2 e^{-\nu} \right] e^\lambda \bigg\},
\end{split}\\
    \frac{1}{\Delta} &= [(l - 1) (l + 2) - 2 (r \omega)^2 e^{-\nu}] [l (l + 1) - 2 (r \omega)^2 e^{-\nu}] e^\lambda + (r \omega)^2 e^{-\nu} \left( 2 - r \frac{d\nu}{dr} \right)^2.
\end{align}
We also know that $\lambda = - \nu$ in vacuum.

\subsection{Static tide}

While  Eq.~\eqref{eq:Exterior} is admittedly more complicated than the equations that are more commonly solved in the stellar exterior, the solution is more immediately linked to the tidal problem. As an illustration, consider the  problem  of static tides \cite{2008ApJ...677.1216H,2009PhRvD..80h4018B}.  This problem corresponds to letting $\omega \rightarrow 0$; in essence, assuming that the tidal driving is slow compared to the induced dynamics. In this regime, Eq.~\eqref{eq:Exterior} reduces to
\begin{equation}
    r^2 \frac{d^2 H_0}{dr^2} + \left( 2 - r \frac{d\lambda}{dr} \right) r \frac{dH_0}{dr} - \left[ l (l + 1) e^{\lambda} + r^2 \left( \frac{d\lambda}{dr} \right)^2 \right] H_0 = 0.
    \label{static}
\end{equation}
Conveniently, this equation admits the general solution  \cite{2008ApJ...677.1216H}
\begin{equation}
    H_0(r) = A Q_l^2(r / M - 1) + B P_l^2(r / M - 1),
    \label{eq:ExteriorStatic}
\end{equation}
where $P_l^2$ and $Q_l^2$ are the associated Legendre polynomials of the first and second kind (of degree $l$), respectively. 
It is worth noting that, due to the spherical symmetry of the background star the perturbation problem is degenerate in the azimuthal index $m$ associated with the decomposition in spherical harmonics $Y_l^m$. This also means that the mode solutions are independent of $m$. 

As we are approaching the tidal problem at the proof-of-principle level, we will focus most of the discussion on the specific case of the quadrupole ($l=2$) problem. In a similar vein, while we aim to demonstrate that the proposed strategy allows us to solve the problem for realistic matter models, we will only provide results for a single equation of state. Specifically, we have chosen the BSk22 model from \cite{2013PhRvC..88f1302G,2018MNRAS.481.2994P,BSkGR}. This is a convenient choice because we can then make direct comparisons with the Cowling approximation results from \cite{2025MNRAS.536.1967C}. Moreover, we will only consider a single stellar model with the canonical mass of $M = 1.4M_\odot$ and a radius of $R=13.03$~km. Once we demonstrate that the methodology is robust one should obviously consider a range of alternative models but we are not overly concerned about this at the present time.

While the solution  \eqref{eq:ExteriorStatic} is valid throughout the exterior spacetime, let us focus our attention on the weak field near zone. 
For the $l=2$ case we then have,  
from the Digital Library for Mathematical Functions\footnote{dlmf.nist.gov},
\begin{equation}
    P_2^2(r/M-1) =  3 \left({r\over M}\right)^2 \left( 1 - {2M\over r} \right),
\end{equation}
along with 
\begin{equation}
    Q_2^2(r/M-1) = \left(\frac{r}{M}\right)^2\left(1-\frac{2M}{r}\right)\left[\frac{M(r-M)\left(2M^2+6Mr-r^2\right)}{r^2\left(r-2M\right)^2}+\frac{3}{2}\ln\left(\frac{r}{r-2M}\right)\right].
    \label{LegendreQ}
\end{equation} 
It follows that, in the weak-field regime, where $M / r \ll 1$, the quadrupole solution to \eqref{static} takes the form 
\begin{equation}
    H_0(r) \approx A\left\{ \frac{8}{5} \left( \frac{M}{r} \right)^3 + \mathcal O[(M / r)^4] \right\} + B \left\{  3\left( \frac{M}{r} \right)^{-2} + \mathcal O[(M / r)^{-1}] \right\}.
    \label{eq:ExteriorStaticl2Mr}
\end{equation}
It is worth noting that the two amplitudes $A$ and $B$ are dimensionless.

We can now identify the decaying term $(A)$ with the stellar response and the growing term $(B)$ with the external tidal field, precisely as we did in the Newtonian case. In fact, in the Newtonian limit we have
\begin{equation}
    e^\nu H_0 \approx 2 (\delta \Phi_l + \chi_l),
\end{equation}
and a comparison with Eq.~\eqref{eq:ExteriorStaticl2Mr} leads to
\begin{equation}
    \delta \Phi_2 = \frac{4}{5} A \left( \frac{M}{r} \right)^3, \qquad \chi_2 = \frac{3}{2} B \left( \frac{M}{r} \right)^{-2}.
\end{equation}
Hence, recalling Eq.~\eqref{kldef}, we infer that
\begin{equation}
    k_2 = {4\over 15} \left({M\over R}\right)^5 \frac{A}{B} .
    \label{kratio}
\end{equation}
This reproduces Eq.~(22) from Ref.~\cite{2008ApJ...677.1216H}. Similarly, it is fairly straightforward to obtain the relativistic version of the matching relation \eqref{eq:Love-alt}. First we introduce $k_2$ in place of one of the two amplitudes, $A$ or $B$, and then combine the expressions for $H_0$ and $dH_0 / dr$ to solve for the remaining one. Since Eq.~\eqref{eq:ExteriorStatic} is valid throughout the exterior spacetime, we can determine $k_2$ by matching at the stellar surface.

\subsection{Near-zone boundary conditions}

Let us now consider the problem of dynamical tides, which naturally involves solving the perturbation problem for finite frequencies. This is not, in itself, very challenging. It is regularly done in order to identify a star's oscillation modes \cite{1998MNRAS.299.1059A}. However, we need to adjust the strategy a little bit. 
In the mode problem the outgoing-wave boundary condition is imposed asymptotically in the wave zone, where the gravitational radiation can be identified. 
This is not what we need for the tidal response. Instead, we have to bring the analysis into the weak-field near zone (as indicated in Figure~\ref{tidefig}). It is only in this region that the matching argument applies. It is obviously not the case that the solution to the binary problem can be obtained by solving for the perturbations of a single star (in the corresponding rest frame), but we can quantify the star's tidal response in the region of spacetime where the tidal driving represents a small perturbation of the star's geometry. 

The general argument may be familiar from recent work within world-line field theory \cite{2021PhRvD.104l4061C,2023PhRvL.130i1403I,2024PhRvD.109f4058S,2024PhRvD.110j3001S,2025arXiv250623176M}, where the tidal response problem is recast in terms of scattering amplitudes. In that context, a key step involves bringing the asymptotic solution into the near-zone to identify the multipole moments of the tidally perturbed body. Our aim here is  similar; we want to work out the  tidal response function which can (equivalently) be thought of as a set of frequency-dependent multipole moments. However, we differ from the (current version of the) world-line field theory argument by explicitly not carrying out a low-frequency expansion. This also makes our calculation distinct from the recent efforts presented in \cite{2024PhRvD.109f4004P} and \cite{2024PhRvD.110d4041R}. We should, however, highlight the fact that our analysis does not provide a complete tidal model. This would require linking the driving frequency $\omega$ to the orbital evolution and we are not taking steps in this direction here. This means, for example, that our results will not demonstrate the anticipated (from the Newtonian analysis) dependence of the tidal response on the $m$-multipoles. This needs to be addressed by future work.

Based on the fact that Eq.~\eqref{eq:ExteriorStatic} provides the general solution for the static tide, which should represent the $\omega\to 0$ limit of the dynamical response, and noting that we want to carry out the matching in the weak-field near zone, we follow \cite{1997PhRvD..56.2118L} and take as starting point the Ansatz 
\begin{equation}
    H_0(r, \omega) = A[1+\alpha (r\omega)^2 ] Q_2^2(r / M - 1) + B[1+\beta (r\omega)^2 ] P_2^2(r / M - 1),
    \label{version1}
\end{equation}
In essence, we express the solution as a double expansion in terms of $M/r \ll 1$ (weak field) and $\omega r\ll 1$ (near zone). The latter is particularly important and it is worth noting that, while the near zone condition has partial overlap with a low-frequency expansion, the two are not identical. 

Substituting into \eqref{eq:Exterior}, we find that the solution to order 
$(r\omega)^2$ requires 
\begin{equation}
    \alpha= -{1\over
    2} + \mathcal O\left( {M\over r}\right) ,
\end{equation}
and
\begin{equation}
    \beta = -{11\over 42} - {107\over 63} {M\over r} +\mathcal O\left[(M/r)^2 \right].
\end{equation}
The first of these agrees with Eq.~(28) from \cite{1997PhRvD..56.2118L}. We find that, while it is sufficient to use a constant value for $\alpha$,  we need to go beyond the leading order in $\beta$.  If we only keep the constant contribution to $\beta$ then the results, e.g. in Figure~\ref{fig:BSKk2}, would have errors at the level of a few percent. We could improve on our analysis by adapting the solution discussed in \cite{2024PhRvD.109j4064H}, but the level of accuracy of our calculation is already beyond the level that can be expected from current and future gravitational-wave observations \cite{2022PhRvD.105l3032W,2025PhRvD.112f3020K}. Nevertheless, future work should include higher order terms to further suppress systematic errors. For the present demonstration, the terms we have included are sufficient.

\subsection{The tidal response function}

At this point, we have prepared the ground for the final step required to complete the matching argument and determine the tidal response function for a relativistic star. We only need to connect the exterior solution from \eqref{version1} to the solution to the interior fluid perturbation problem.  

Matching the two solutions at the surface of the star, we have
\begin{equation}
    H_0(R, \omega) = A \left[1+\alpha (R\omega)^2 \right] Q_2^2(R / M - 1) + B\left[1+\beta (R\omega)^2 \right] P_2^2(R / M - 1),
\end{equation}
and
\begin{multline}
    H_0'(R, \omega) = A\left\{ 2\alpha \omega^2 R  Q_2^2(R / M - 1) +  \left[1+\alpha (R\omega)^2 \right] \left. {dQ_2^2 \over dr} \right|_{R / M - 1}\right\}   \\
    + B \left\{ 2\beta \omega^2 R P_2^2(R / M - 1) +   \left[1+\beta (R\omega)^2 \right] \left. {dP_2^2\over dr} \right|_{R / M - 1} \right\} .
\end{multline}
Once we invert these relations to determine the two amplitudes $A$ and $B$  the tidal response is obtained from \eqref{kratio}. The construction of the near-zone boundary condition guarantees that the result limits to the static Love number as $\omega\to 0$.

Schematically, we write the two equations as 
\begin{eqnarray}
    H_0 &=& a_1 A + a_2 B ,\\
    H_0'  &=& a_3 A + a_4 B,
\end{eqnarray}
to get
\begin{equation}
    a_3 H_0 - a_1 H_0' = (a_2 a_3- a_1 a_4)  B  \Longrightarrow B = - {1\over a_1a_4 - a_3 a_2} \left[ a_3 H_0 - a_1 H_0'\right].
\end{equation}

From this we see that the star's oscillation modes correspond 
\begin{equation}
    a_3 H_0 - a_1 H_0' = 0 ,
\end{equation}
or 
\begin{equation}
  \left\{ 2\alpha \omega^2 R  Q_2^2(R / M - 1) +  \left[1+\alpha (R\omega)^2 \right] \left. {dQ_2^2 \over dr} \right|_{R / M - 1}\right\} H_0(R,\omega) -  \left[1+\alpha (R\omega)^2 \right] Q_2^2(R / M - 1) H_0'(R,\omega) = 0  .
      \label{modelcon1}
\end{equation}
Strictly speaking this condition only implies that the solution decays away from the star. It is not (yet) a statement of purely outgoing waves at asymptotic distances. As a result, the mode frequencies one obtains from \eqref{modelcon1} will be real valued. 
We will return to the issue of outgoing waves later. 

We also need 
\begin{equation}
a_4 H_0 - a_2 H_0' = ( a_1 a_4- a_2 a_3 ) A \Longrightarrow A= {1\over a_1a_4 - a_3 a_2} \left[ a_4 H_0 - a_2 H_0'\right].
\end{equation}
The general expression for the tidal response, obtained from matching at the stellar surface, then becomes 
\begin{equation}
    k_2 = - {4\over 15} \left({M\over R}\right)^5 { a_4 H_0 - a_2 H_0' \over  a_3 H_0 - a_1 H_0' }.
    \label{k2match}
\end{equation}

\begin{figure}
\centering\includegraphics[width=0.6\textwidth]{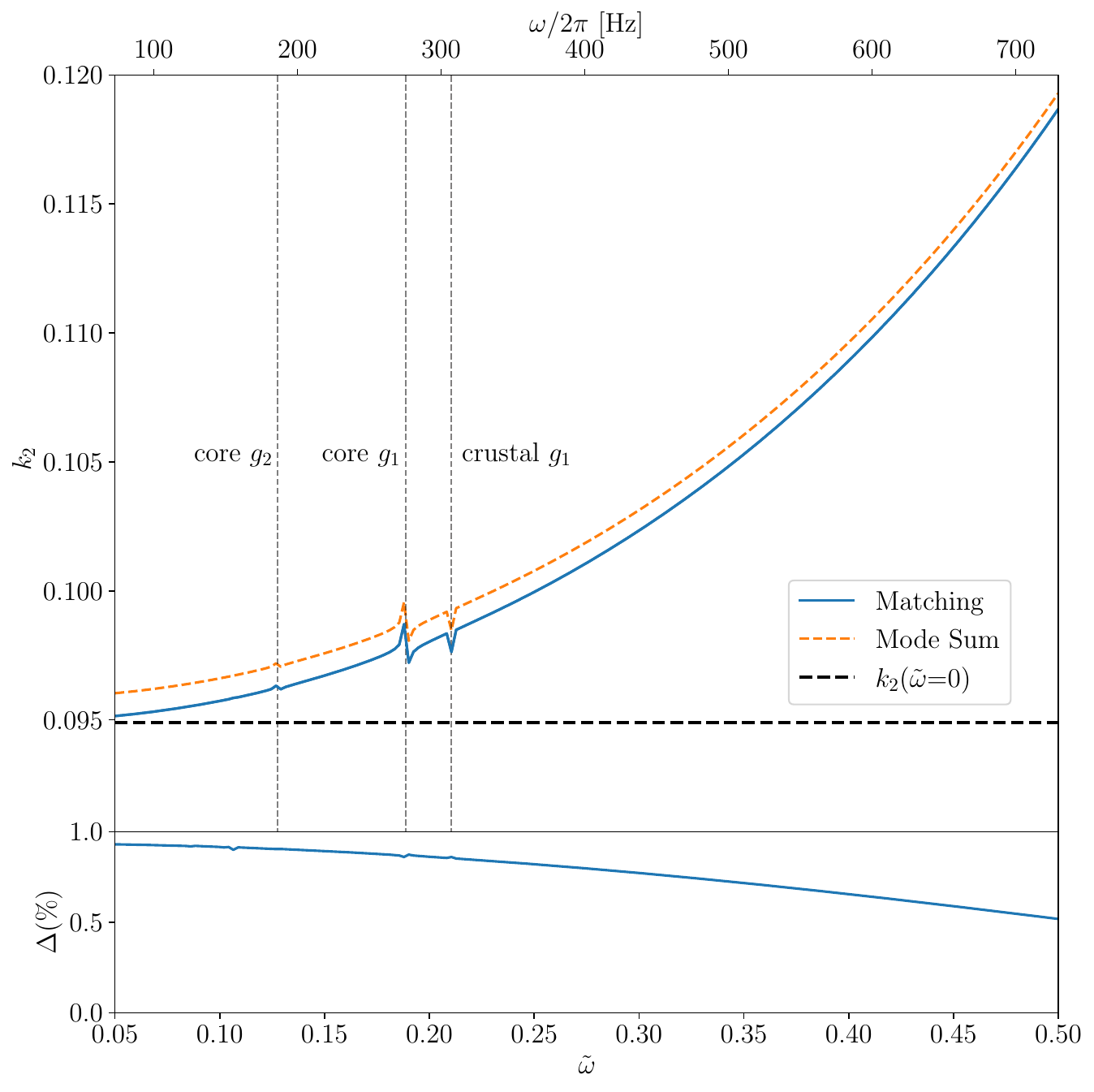}
    \caption{ Top panel: The effective Love number $k_2(\omega)$ obtained from the matching relation \eqref{version1} for a $1.4M_\odot$ neutron star and the BSk22 equation of state (solid blue curve). The static result, $k_2\approx0.0949$ obtained following the standard calculation from \cite{2008ApJ...677.1216H}, is indicated (horizontal dashed line) as are the first two core g-modes and a crustal g-mode (vertical dashed lines). Also shown (as a dashed orange curve) is the result obtained from the inferred mode sum. Bottom panel: The relative (percentage) error, $\Delta$, in the inferred mode sum compared to the result for $k_2$ obtained from matching at the star's surface. With a typical error below the 1\% level---likely indicative of the numerical accuracy of the implementation---it is evident that the analysis leads to robust results for the $\mathcal A_n$ amplitudes. The result also shows that, for all practical purposes one can safely replace the matching calculation with a sum over the modes. This  agrees with the intuition gleaned from the Newtonian case and neatly circumvents the non-Hermitian nature of the relativistic problem.}
    \label{fig:BSKk2}
\end{figure}

In order to gain confidence in the approach, we have implemented the near-zone condition and calculated oscillation modes for the $1.4M_\odot$ BSk22 model. This model includes composition gradients, and hence supports the existence of low frequency g-modes. A sample of results are provided in Table~\ref{ModeTable} and compared to a fully relativistic  calculation (based on the implementation for the internal problem and the usual asymptotic outgoing-wave conditions from \cite{2025PhRvD.111h3024G}). The results show that the near-zone boundary condition \eqref{modelcon1} enables an accurate calculation of the real part of the mode frequency. This accords well with the discussion in \cite{1997PhRvD..56.2118L}. The numerical results in Table~\ref{ModeTable} should be indicative of the precision with which we can obtain the tidal response function from the matching argument. A typical frequency dependent tidal response obtained from \eqref{k2match} is provided in Figure~\ref{fig:BSKk2}, which covers the low-frequency region and the first couple of g-modes for the BSk22 model.

\renewcommand{\arraystretch}{1.3}
\begin{table}[ht]
\normalsize
\centering
\caption{Relativistic mode results for the $M=1.4M_\odot$ stellar model and the BSk22 equation of state. We provide data for the f-mode, the first couple of g-modes and a g-mode specifically associated with the crust region (see \cite{2025MNRAS.536.1967C} for comments on this specific mode) within the relativistic Cowling approximation as well as from the two versions of the near-zone boundary condition, Eq.~\eqref{version1} and Eq.~\eqref{H0final}. A comparison to the fully relativistic calculation (described in \cite{2025PhRvD.111h3024G}) provides confidence in the near-zone approximation. We also see that, while the simpler condition from Eq.~\eqref{version1} accurately determines the real part of the f-mode frequency, we need the improved condition in Eq.~\eqref{H0final} if we also want to calculate the imaginary part. The results are in complete agreement with the expectations from \cite{1997PhRvD..56.2118L}.}
\begin{tabular}{|c|c|c|c|c|}
\hline
\multicolumn{1}{|c}{\multirow{2}{*}{Mode}} & \multicolumn{4}{|c|}{\text{Re\,}$(\tilde \omega_n)$}\\
\cline{2-5}
\multicolumn{1}{|c|}{} & Cowling & GR & Eq.\,\eqref{version1} & Eq.\,\eqref{H0final}
\\
\hline

                                $f$ & 1.4224 &	1.1272 &	1.1274&	1.1274 \\
                                \text{crustal} $g_1$ &0.2111 &	0.2105&	0.2105 &	0.2105 \\
                                \text{core }$g_1$ & 0.1822	&0.1888	&0.1888&0.1888\\
                               \text{core }$g_2$ & 0.1263	&0.1278&	0.1278&	0.1278\\

            \hline
\multicolumn{1}{|c}{} & \multicolumn{4}{|c|}{\text{Im\,}$(\tilde \omega_n)$}\\
\hline

                                $f$ & -- &	1.26E-05 & --	&1.13E-05\\
\hline
\end{tabular}
\label{ModeTable}
\end{table}

As in the Newtonian case, we can turn the matching result into a mode sum by Taylor expansion near each mode frequency. However, this now involves a leap of faith. In the absence of a demonstration that the relativistic modes form a useful basis---noting that the closest result we have in this direction is obtained from matching to a post-Newtonian tidal environment \cite{2025arXiv250710693A}---the construction is phenomenological. Having said that, the argument provides a simple representation of the tidal response that would be easy to use in applications.

Close to a given mode frequency, we can Taylor expand to get 
\begin{equation}
 a_3 H_0 - a_1 H_0' \approx (\omega - \omega_n) \partial_\omega \left[ a_3 H_0 - a_1 H_0' \right]_{\omega=\omega_n}.
\end{equation}
In general, we need to execute a bit of caution here. The relativistic modes have complex frequencies so if we adapt the Newtonian calculation and evaluate the frequency derivative on the real $\omega$ axis then we incur numerical errors. These errors should be small enough that they can be ignored for low-frequency gravity modes, which are known to be extremely slowly damped by gravitational-wave emission. The situation for the fundamental f-mode is not so clear.  However, 
 for a mode with frequency $\omega_n = \sigma_n + i \gamma_n$ (recall that we assume all perturbations have a harmonic time dependence $e^{i\omega t}$) we would have
\begin{equation}
   \partial_\omega \left[ a_3 H_0 - a_1 H_0' \right]_{\omega=\omega_n}\approx {i\over \gamma_n} (a_3 H_0 - a_1 H_0')_{\omega=\sigma_n}  ,
\end{equation}
as long as the imaginary part is small enough. This relation provides useful insight into the precision required for the calculation. Specifically, the accuracy to which we are able to determine $\gamma_n$ directly impacts on the numerical result for the mode contribution to the tidal response.

In order to express the tidal response as a mode sum, we also need to keep in mind the expected symmetry of the mode solutions. Given that the frequency only enters as $\omega^2$ in the perturbation equations, the modes appear as pairs with frequencies $\pm \sigma_n +i\gamma_n$. This means that we should have (using the same dimensionless frequency as in the Newtonian problem)
\begin{equation}
    k_{l} = - \frac{2 \pi }{2 l + 1 }
            \sum_{n} {\mathcal A_n  \over 
            (\tilde \omega + \tilde \sigma_n-i \tilde \gamma_n)(\tilde \omega- \tilde \sigma_n-i \tilde \gamma_n)} ,
\end{equation}
where $\mathcal A_n$ replaces $\tilde Q_n^2$ from the Newtonian calculation.
This leads to 
\begin{equation}
\mathcal A_n =  -  {2l+1 \over \pi} \tilde \sigma_n \lim_{\tilde \omega \to \tilde \omega_n} ( \tilde \omega -\tilde \omega_n) k_l(\tilde\omega) =  - {2l+1 \over \pi} (\mathrm{Re}\,\tilde \omega_n) \mathrm{Res}\, k_l(\tilde \omega_n) ,
\label{Aninfer}
\end{equation}
for which we need
\begin{equation}
\mathrm{Res}\, k_l(\tilde \omega_n)  = - {4\over 15} \left({M\over R}\right)^5 \left[ a_4 H_0 - a_2 H_0' \right] \left[ \left. {\partial \left( a_3 H_0 - a_1 H_0' \right) \over \partial\omega} \right|_{\omega=\omega_n} \right]^{-1}.
\end{equation}
In practice, this relation is only useful for modes for which $\tilde \gamma_n$ is non-negligible, like the fundamental mode. For low-frequency g-modes the calculation may equally well be carried out assuming that the frequency is real.

A sample of numerical results for our chosen BSk22 model are provided in Table~\ref{GROverlapTable} (with the required residues simply obtained by linear interpolation at each resonance). For comparison we have included the corresponding results obtained within the relativistic Cowling approximation \cite{2025MNRAS.536.1967C}. While we expect the Cowling approximation to be fairly accurate for g-modes, we know the error in the f-mode frequency will be of order $15-20\%$. With this in mind, one might anticipate the amplitude $\mathcal A_n$ obtained for g-modes from the matching argument to be similar to the overlap integral $\tilde Q_n^2$ from the Cowling calculations. However, the numerical results suggest that the tidal excitation of all modes is significantly weaker in the relativistic case. The amplitude $\mathcal A_n$ is about a factor of 3 smaller than the result for $\tilde Q_n^2$ obtained within the relativistic Cowling approximation. The $g_1$-mode excitation is similarly suppressed by a factor of a few, while the $g_2$ mode amplitude is more than an order of magnitude smaller than the Cowling result. The behaviour is evidently mode dependent and we will need a broader survey of stellar models to establish a clear pattern.  

\renewcommand{\arraystretch}{1.3}
\begin{table}[ht]
\normalsize
\centering
\caption{Comparing the overlap integral $\tilde Q_n^2$ for the modes obtained within the relativistic  Cowling approximation and the inferred  (real valued) $\mathcal A_n$ from \eqref{Aninfer}. The results indicate that the tidal excitation of all modes is significantly weaker in the relativistic case.}
\begin{tabular}{|c|c|c|}
\hline
Mode & $\tilde Q_n^2$ & $\mathcal{A}_n$ 
\\
\hline

                                $f$ & 2.83E-01 &	9.69E-02  \\
                                \text{core} $g_1$ & 8.40E-07&2.48E-07 \\
                               \text{core }$g_2$ & 2.80E-07&	1.91E-08\\
                               \text{crustal} $g_1$ & 3.30E-06 & 8.29E-09 \\

            \hline
\end{tabular}
\label{GROverlapTable}
\end{table}

It is also useful to note that the static limit provides a sanity check of the result. Specifically, we know that the mode sum should limit to the static tide in the zero-frequency limit. Assuming that the f-mode provides the dominant contribution, which is evident from the numerical results, we should then have 
\begin{equation}
    k_2 (\omega=0) \approx {2\pi \over 5} {\mathcal A_f \over \tilde \sigma_f^2 }\Longrightarrow \mathcal A_f \approx {5\over2\pi} \tilde \sigma_f^2  k_2 (\omega=0).
\end{equation}
Putting numbers in, with $k_2\approx 0.0949$ and $\tilde \sigma_n =  1.1274$ we would expect $\mathcal A_f\approx 0.096$, in perfect agreement with the result from Table~\ref{GROverlapTable}. The argument also suggests that $\mathcal A_n$ should be a real-valued quantity even when we account for the gravitational-wave damping of the modes. 

Knowing that the individual $\mathcal A_n$ amplitudes should be real, it is easy to show that, in general,
\begin{equation}
    \mathrm{Re}\ k_{l} 
            = - \frac{ 2\pi }{2 l + 1 }
            \sum_{n} \mathcal A_n   { \tilde \omega^2 - (\tilde \sigma_n^2 + \tilde \gamma_n^2)  \over 
            \left[ \tilde \omega^2 - (\tilde \sigma_n^2 + \tilde \gamma_n^2)\right]^2 +4 (\tilde \omega \tilde \gamma_n)^2 } ,
\end{equation}
and
\begin{equation}
     \mathrm{Im}\ k_{l} 
            = - \frac{ 2\pi }{2 l + 1 }
            \sum_{n} \mathcal A_n   { 2  \tilde \omega \tilde \gamma_n \over 
            \left[ \tilde \omega^2 - (\tilde \sigma_n^2 + \tilde \gamma_n^2)\right]^2 +4 (\tilde \omega \tilde \gamma_n)^2 }.
            \label{Imk}
\end{equation}
It is worth noting that for a complex mode frequency the mode contribution to the real part of $k_l$ vanishes at resonance, which means that the  tidal response is dominated by the imaginary part of $k_l$ close to resonance. Given that the imaginary part of $k_l$ is directly linked to the tidal lag and the torque acting on the star \cite{2024PhRvD.110b4039Y}, this is a relevant observation.

Finally, it is useful to compare the frequency dependent tidal response and the inferred mode sum. This comparison, which is provided in Figure~\ref{fig:BSKk2}, suggests that the phenomenological mode sum is accurate to better than 1\% across the relevant frequency range. This, first of all, indicates that the analysis used to determine the $\mathcal A_n$ amplitude is robust. Moreover, given that we do not expect current or future observations to be able to probe the tidal response at this level of precision \cite{2022PhRvD.105l3032W,2025PhRvD.112f3020K}, we may  safely replace the matching calculation with the effective sum over the modes in relevant applications. 

\subsection{Adding gravitational-wave damping}

The proposed matching strategy for the tidal response of a relativistic star relies on a near-zone boundary condition for the perturbation problem and we argued that a comparison of oscillation mode frequencies obtained from this approach to the fully relativistic mode calculation provide confidence in the results. However,  the calculation obviously did not account for the anticipated imaginary part of each mode frequency. Of course, we know from the original analysis of the near-zone boundary conditions \cite{1997PhRvD..56.2118L}, that we need to take an extra step to improve the calculation. Whether this step is required for the tidal response calculation remains to be established. We know from \eqref{Imk} that the imaginary part determines the behaviour near resonance, but other factors (like the orbital evolution) will also come into play \cite{2025MNRAS.542.1375P}. Regardless, it is clearly desirable to improve on the calculation to remove controllable errors.

Let us recall the argument that led to the mode condition \eqref{modelcon1}: First we solved the static problem (in terms of the associated Legendre functions) and then we added the leading order near-zone frequency corrections. In essence, the starting point was a solution that is accurate to all orders in $M/r$ for $\omega =0$. We could have introduced the approximations in the reverse order. That is, first made the near-zone approximation $\omega r \ll 1$ and then added curvature corrections. This calculation was first done in a seminal paper by Thorne \cite{1969ApJ...158..997T}. He started from  Eq.~\eqref{eq:Exterior} in the weak-field limit,  $M / r \ll 1$. We then have
\begin{equation}
    r^2 \frac{d^2 H_0}{dr^2} + (2 + r \eta_1) r \frac{dH_0}{dr} + [(r \omega)^2 - l (l + 1) + r^2 \eta_2] H_0 + \mathcal O(M / r) = 0,
    \label{eq:ExteriorMr}
\end{equation}
where
\begin{align}
    r \eta_1 &= 4 \{ (l - 1) (l + 2) \Delta [l (l + 1) - 2 (r \omega)^2] - 1 \} + \mathcal O(M / r), \\
    r^2 \eta_2 &= 16 (r \omega)^4 \Delta + \mathcal O(M / r),
\end{align}
and
\begin{equation}
    \frac{1}{\Delta} = [(l - 1) (l + 2) - 2 (r \omega)^2] [l (l + 1) - 2 (r \omega)^2] + 4 (r \omega)^2 + \mathcal O(M / r).
\end{equation}
This equation can also be solved analytically, although now the solutions are expressed in terms of spherical Bessel functions ($j_l$ and $y_l$). Specifically, we have
\begin{equation}
    H_0 =  r \omega \left[ C \frac{d}{d (r \omega)} j_l(r \omega) + D \frac{d}{d (r \omega)} y_l(r \omega) \right] + \left[ 1 + \frac{1}{2} l (l + 1) - (r \omega)^2 \right] [C j_l(r \omega) + D y_l(r \omega)] ,
    \label{eq: SmallMR}
\end{equation}
where $C$ and $D$ are arbitrary constants.
Well away from the star (in the wave zone $r \omega \gg 1$) it is easy to show that
\begin{equation}
    C j_l(r \omega) + D y_l(r \omega) \approx - \frac{1}{2 r \omega} \left[ (D + i C) e^{i (r \omega - l \pi / 2)} + (D - i C) e^{- i (r \omega - l \pi / 2)} \right] .
    \label{eq: AsymptoticWaveZone}
\end{equation}
We see that---keeping in mind the $e^{i\omega t}$ time dependence---a solution with no ingoing radiation (representing an oscillation mode for the star) corresponds to $D = - i C$. This identification is similar in spirit to setting $\hat{\chi}_l = 0$ in Eq.~\eqref{eq:Relation} in the Newtonian problem.

Next, we consider the solution in the near zone, where $r \omega \ll 1$. We then have
\begin{equation}
    C j_l(r \omega) + D y_l(r \omega) \approx C \frac{(r \omega)^l}{(2 l + 1)!!} \{ 1 + \mathcal O[(r \omega)^2] \} - D \frac{(2 l - 1)!!}{(r \omega)^{l + 1}} \{ 1 + \mathcal O[(r \omega)^2] \}.
\end{equation}
These relations allow us to bring the outgoing-wave boundary condition into the weak-field near zone, albeit in an approximate fashion \cite{1997PhRvD..56.2118L}. 

In order to complete the sequence of approximations, we want to account for the anticipated imaginary part from \eqref{eq: SmallMR}, in the limit $M / r \ll 1$, while at the same time retaining the static tide limit from \eqref{version1}. As argued by Lindblom et al \cite{1997PhRvD..56.2118L}, we can do this by matching the leading order terms in the expansion of each of the two solutions. This leads to the combined expression 
\begin{multline}
    H_0(r,\omega) = A\bigg\{[1+\alpha (r\omega)^2 ] Q_l^2(r / M - 1) -iK_l \left[ r \omega \frac{d}{d (r \omega)} + 1 + \frac{1}{2} l (l + 1) - (r \omega)^2 \right]  j_l(r \omega) \bigg\} \\
    + B[1+\beta (r\omega)^2 ] P_l^2(r / M - 1),
    \label{H0final}
\end{multline}
with
\begin{equation}
    K_l=\frac{2(l+2)!(M \omega)^{l+1}}{l(l-1)(2l-1)!!(2l+1)!!},
\end{equation}
In the limit $(r \omega)^2 \ll 1$ and $\omega \rightarrow 0$ the solution from \eqref{H0final} reduces to \eqref{eq:ExteriorStatic}. Moreover, the constant $K_l$ ensures that the imaginary part of $H_0(r)$ reduces to Eq.~(27) in \citet{1997PhRvD..56.2118L} in the appropriate  $(r \omega)^2 \ll 1$ and $M / r \ll 1$ limit. 

It is worth noting that we only adjust for the wave behaviour in the first term in \eqref{version1}. This makes sense because this is the solution that is associated with the outgoing gravitational waves in the distant wave zone, as indicated in Figure~\ref{tidefig}.  The other solution is identified with the tidal driving and does not involve wave-like behaviour far away from the system. At least not until we consider the ``outer problem'' and the connection between the tidal response and the orbital evolution.

The numerical results for the BSk22 model confirms that the revised near-zone condition \eqref{H0final} leads to a fairly accurate result for the imaginary part of the f-mode frequency, see Table~\ref{ModeTable}. Furthermore, it would be straightforward to extend the calculation of the tidal response to account for this. We have not done so here simply because the results provided in Figure~\ref{fig:BSKk2} are already at a satisfactory level of precision.

\subsection{Translating to the scattering problem}

A growing body of recent work on the tidal problem draws on strategies adapted from effective field theory. In particular, it is common to map the tidal response problem onto a scattering problem, where the key information is contained in the asymptotic behaviour of the perturbations (in the wave zone) \cite{2021PhRvD.104l4061C,2023PhRvL.130i1403I,2024PhRvD.109f4058S,2024PhRvD.110j3001S,2025arXiv250623176M}. This then allows an extraction of the frequency dependent tidal information. From a conceptual perspective, keeping in mind the schematic illustration in Figure~\ref{tidefig}, the attitude we have taken here---specifically, matching to the tidal interaction in the weak-field near zone---may seem more natural, especially since one can see how one may expend the calculation to numerical simulations (using the world-tube argument explored in, for example, \cite{2023PhRvD.108b4041W,2024PhRvD.110h4023W}).
At the same time, it is clearly beneficial to make a direct connection to the scattering problem. A combination of results may  facilitate progress on the orbital evolution evolution problem and the generation of the all-important gravitational waveform models.

Motivated by these arguments, let us demonstrate (as far as we are aware, for the first time---although see \cite{2024PhRvD.109f4058S} for the corresponding low-frequency near-far factorisation for black holes) how the near-zone matching calculation can be connected to the scattering problem. This turns out to be straightforward. First of all, from the steps we have already taken we see that the ingoing solution can be approximated by 
\begin{equation}
    H_0 \approx   Q_2^l + {8 i \over 15} (\omega M)^3   \left[   r \omega \frac{d}{d (r \omega)} + 1 + \frac{1}{2} l (l + 1) - (r \omega)^2 \right] j_l (\omega r).
\end{equation}
This then implies that the near-zone solution to the scattering problem should take the form
\begin{multline}
    H_0(r,\omega) \approx C_\mathrm{out}\bigg\{ Q_l^2(r / M - 1) -iK_l \left[ r \omega \frac{d}{d (r \omega)} + 1 + \frac{1}{2} l (l + 1) - (r \omega)^2 \right]  j_l(r \omega) \bigg\} \\
    + C_\mathrm{in}\bigg\{ Q_l^2(r / M - 1) + iK_l \left[ r \omega \frac{d}{d (r \omega)} + 1 + \frac{1}{2} l (l + 1) - (r \omega)^2 \right]  j_l(r \omega)\bigg\},
    \label{H0scat}
\end{multline}
with $C_\mathrm{out}$ and $C_\mathrm{in}$ representing the  amplitudes of out- and ingoing-wave solutions in the near zone, respectively. The amplitudes are proportional to the more commonly discussed asymptotic wave amplitudes, $A_\mathrm{out}$ and $A_\mathrm{in}$, defined in the wave zone. We provide a few comments on this relation in Appendix~\ref{app:scattering}.

At first sight, the solution in \eqref{H0scat} seems quite different from the version from \eqref{H0final} that we employed in the tidal matching argument. However, the connection between the two is easily established once we note that, for a given frequency $\omega$, the solution to the interior perturbation problem remains unchanged. As a direct demonstration, let us keep only the leading order terms in each solution (for $l=2$). From \eqref{H0final} we then have, collecting terms in the $M/r$ expansion
\begin{equation}
H_0 \approx A \left( {M\over r} \right)^3  +  \left\{ 3B   - {2iA\over 15} (\omega M)^5  \right\} \left( {r\over M} \right)^2,
\end{equation}
while the  scattering solution \eqref{H0scat} leads to
\begin{equation}
H_0 \approx (C_\mathrm{out}+C_\mathrm{in}) \left( {M\over r} \right)^3 + {2i\over 15} (C_\mathrm{out}-C_\mathrm{in}) (\omega M)^5  \left( {r\over M} \right)^2.
\end{equation}
Identifying the coefficients we find that  
\begin{equation}
    C_\mathrm{out} =  A + {45i\over 4 (\omega M)^5} B,
\end{equation}
and 
\begin{equation}
    C_\mathrm{in} = - {45i\over 4 (\omega M)^5} B,
\end{equation}
The latter relation nicely demonstrated that the mode solutions will be the same in the two case, given that  $C_\mathrm{in}=0\Longrightarrow B=0$.
The translation shows that we can alternatively work out the tidal response from \eqref{kratio} or
\begin{equation}
    k_2 = - {3i \over (\omega R)^5} \left[ 1 + {C_\mathrm{out} \over C_\mathrm{in} } \right].
\end{equation}
In this case the $\omega\to 0$ limit clearly must be approached with some caution but it turns out that terms in the expression conspire to cancel to make the expression regular (and identical to the one we used in our calculations). It is also worth commenting on the fact that the behaviour near resonance will be dominated by the singularity associated with the vanishing of $C_\mathrm{in}$. Again, the calculation will involve a residue argument \cite{2025arXiv250623176M}.
Finally,
in order to make it practically useful, one should obviously take this argument beyond the leading order demonstration we have outlined here. However, it is now clear how this calculation can be done which is a useful step in the right direction.

\section{Concluding remarks}

We have develop a fully relativistic approach for determining the frequency-dependent tidal response of a compact star. The strategy involves matching the solution for the linearised fluid dynamics in the star's interior to the spacetime perturbations in the near zone surrounding the body, along with an identification of the tidal driving and the star's response. We have provided an explicit demonstration of the fact that this identification is exact in Newtonian gravity and argue that the approach remains robust also in the relativistic case. Notably, the calculation does not involve a sum over the star's quasinormal modes and hence circumvents one of the obstacles that have held up the development of models for relativistic tides \cite{2024PhRvD.109f4004P}. Our numerical results, while admittedly at a proof-of-principle level, illustrate the method for a realistic matter model from the BSk family, including composition stratification leading to the presence of low-frequency gravity modes. We believe this represents the first set of fully relativistic dynamical tide results for a realistic neutron star equation of state. 

The results for the tidal response rely on an implementation of near-zone boundary conditions for the perturbed spacetime metric. This approach was first advocated by Lindblom \emph{et al.} some time ago \cite{1997PhRvD..56.2118L}. We have revisited the argument and, for the first time, confirmed the accuracy of the strategy for a realistic neutron star model. As an extension of the near-zone argument we have also made contact with the body of work that translates the problem into that of wave scattering, e.g. \cite{2024PhRvD.110j3001S}, with the key information encoded in the asymptotic amplitudes (see, for example, the recent discussion in \cite{2025arXiv250623176M}). The explicit relation between the two approaches should enable direct comparison of  results, as required to establish confidence in the respective methods.

On the note of comparison, we have only made direct reference to previous results obtained within the relativistic Cowling approximation \cite{2025MNRAS.536.1967C}. This comparison is more indicative than satisfactory, because it is well known that the fully relativistic fundamental modes differs from the modes obtained in the Cowling approach by 15-20\%. However, at the present time no other direct comparison---for realistic neutron star matter---is available to us. This should hopefully change in the near future once the mode-sum strategy outlined in \cite{2025arXiv250710693A} is completed. It would seem highly relevant to compare results from our matching argument to this new mode-sum formulation. Both approaches involve approximations and if one could demonstrate that they agree then this would provide mutual confidence.

As there are a number of ways in which the matching strategy can be developed and improved, let us close with a few general remarks. First, a great advantage of the matching argument is that one can readily deal with any  internal structure. As long as we can solve the perturbation equations, we may include an elastic neutron star crust, superfluid components, internal phase transition, dissipation and so on. Given the large body of work aimed at developing mode calculations for realistic neutron star configurations \cite{2015PhRvD..92f3009K,2025PhRvD.111h3024G}
extensions in this direction would come more or less for ``free''. Second, it is clear that the effective mode-sum we obtain through the residue argument lends support for the commonly used Effective One Body models \cite{2016PhRvD..94j4028S}, which build on the assumption that the fluid response can be represented by adding a ``harmonic oscillator'' contribution to the system's Hamiltonian.  This argument relies heavily on the Newtonian mode-sum intuition. The arguments we have provided here lends support for the strategy, which is (at least conceptually) satisfying. Having said that, the obvious weakness of the matching argument is that we do not have access to an underlying equation for the mode amplitudes that would allow us to extend the near resonance solution from, for example, \cite{2024PhRvD.110b4039Y,2025MNRAS.542.1375P} to relativity. 
We have also not established the link to the orbital evolution---as required to, for example, work out an explicit expression for the  ``energy'' associated with the tidal response. In essence, there is still a fair amount of work left to do.

\begin{acknowledgments}
We would, first of all, like to acknowledge helpful discussions with Abishek Hegade, Muddu Saketh and Hang Yu on relevant aspects of the project. In addition, 
N.A. and S.G. gratefully acknowledge support from the STFC via Grant No.~ST/Y00082X/1.
F.G. acknowledges funding from the European Union’s Horizon Europe research and innovation programme under the Marie Sk{\l}odowska-Curie Grant Agreement No.~101151301.
N.A. and F.G thank the Institute for Nuclear Theory at the University of Washington for its kind hospitality and stimulating research environment. 
This research was supported in part by the Institute's U.S. Department of Energy Grant No.~DE-FG02-00ER41132.
\end{acknowledgments}

\appendix

\section{\label{app:scattering}Explicitly linking to the asymptotic scattering amplitudes}

In the main discussion we made a connection between the tidal response and the problem of waves scattered by the star. This relation is useful given ongoing efforts to extract tidal parameters from scattering amplitudes \cite{2021PhRvD.104l4061C,2023PhRvL.130i1403I,2024PhRvD.109f4058S,2024PhRvD.110j3001S,2025arXiv250623176M}. Moreover, noting that our matching approach is not gauge invariant, it makes sense to further explore this link. In this Appendix we take some steps in this direction, admittedly leaving several details for future work.

Let us first recall \eqref{eq: SmallMR}, which leads to
\begin{equation}
    H_0 \approx - (r \omega)^2 [C j_l(r \omega) + D y_l(r \omega)] \approx \frac{\omega r }{2} \left[ (D + i C) e^{i (r \omega - l \pi / 2)} + (D - i C) e^{- i (r \omega - l \pi / 2)} \right] .
    \label{Hasymp}
\end{equation}
in the  wave zone where $\omega r \gg 1$.
With the assumed time dependence $e^{i\omega t}$ we get
\begin{equation}
    H_0(t,r) \approx  \frac{\omega r }{2} \left[ (D + i C) e^{i  \omega  (t+r)  - i l \pi / 2} + (D - i C) e^{ i  \omega (t-r) + l \pi / 2} \right] ,
\end{equation}
and it follows that a solution with no ingoing radiation (representing an oscillation mode for the star) corresponds to $D = - i C$ while a purely ingoing solution follows from $D=iC$. Given this we identify the near-zone amplitudes corresponding to the two degrees of freedom: $C_\mathrm{out} = D-iC$ and $C_\mathrm{in} = D+iC$. We now want to connect these amplitudes to the asymptotic solutions in the wave zone. 

The natural way to execute the translation is to make contact with the standard solution to the vacuum perturbation problem. The relevant solutions then follow from the Zerilli (or Regge-Wheeler) equation, see \cite{2019gwa..book.....A} for context and further details. This involves explicitly gauge-invariant combinations of the metric perturbations. Specifically, we need \cite{1992PhRvD..46.4289K}
\begin{equation}
    Z = {r \over \omega^2 r^3 - (n+1)M} \left[ {nr^2 - 3nMr-3M^2 \over nr+3M} K-re^{-\lambda} H_0\right] , 
\end{equation}
with 
\begin{equation}
    n = {(l-1)(l+2) \over 2},
\end{equation}
and
\begin{equation}
    e^{-\lambda} = 1- {2M\over r}.
\end{equation}
The asymptotic solution then takes the form
\begin{equation}
    Z = A_\mathrm{out} e^{-i\omega r_\ast} + A_\mathrm{in} e^{i\omega r_\ast},
\end{equation}
where we have introduced the tortoise coordinate
\begin{equation}
    r_\ast = r+ 2M \ln \left( {r\over 2M} -1\right) + \mathrm{constant}.
\end{equation}

At large distances, $r\gg M$, we have 
\begin{equation}
    Z \approx  {1 \over \omega^2 r} \left(  K- H_0\right) .
\end{equation}
From the discussion in \cite{1992PhRvD..46.4289K} we know that the perturbation equations lead to the relations
\begin{equation}
    H_0' + {e^{\lambda}(r-3M) \over r^2} K- e^{\lambda} {(r-4M) \over r^2} H_0=  -\left( \omega^2 e^\lambda - {n+1 \over r^2} \right) {1\over i\omega} H_1,
\end{equation}
and 
\begin{equation}
(\omega^2 r^4 e^\lambda-nr^2-Mr+M^2e^\lambda)K+ (nr+3M)r H_0 = {1\over i\omega} [ \omega^2 r^3 -(n+1)M] H_1  .  
\end{equation}
These relations allow us to work out the Zerilli function associated with our near-zone boundary conditions. However, for the moment we are interested in the behaviour far from the star.
To leading order the relations become
\begin{equation}
    H_0' + {1 \over r^2} K- {1 \over r} H_0 \approx -\omega^2 {1\over i\omega} H_1,
\end{equation}
and
\begin{equation}
\omega^2 r^4 K+ nr^2 H_0 \approx {1\over i\omega}  \omega^2 r^3 H_1  .
\end{equation}
Combining these to remove $H_1$, we have
\begin{equation}
    K \approx - {1\over \omega^2 r^2} \left[  r  H_0' +(n-1) H_0 \right] .
\end{equation}
Recalling the asymptotic behaviour from \eqref{Hasymp} we know that 
\begin{equation}
    H_0 \approx  \frac{\omega r }{2} \left[ C_\mathrm{in} e^{i (r \omega - l \pi / 2)} + C_\mathrm{out} e^{- i (r \omega - l \pi / 2)} \right] ,
\end{equation}
which leads to
\begin{equation}
    Z \approx  {1 \over \omega^2 r} \left(  K- H_0\right) \approx  - {1 \over \omega^2 r} H_0 \approx  - \frac{1 }{2 \omega } \left[ C_\mathrm{in} e^{i (r \omega - l \pi / 2)} + C_\mathrm{out} e^{- i (r \omega - l \pi / 2)} \right] .
\end{equation}
From this it follows that
\begin{equation}
    {A_\mathrm{out} \over A_\mathrm{in}} \approx {C_\mathrm{out} \over C_\mathrm{in}}  e^{i \pi l}, 
\end{equation}
which links the asymptotic amplitudes to the ones determined in the near zone. As presented here, the argument does not account for wave-scattering off of the spacetime curvature, which also affects the phase-shift in the amplitude relation. Adding this aspect to the calculation---essentially taking the argument beyond leading order---would be required to explicitly link our results to the work from, for example,  \cite{2025arXiv250623176M}. Such an effort would also provide the grey-body factor required for the near-zone to far-zone relation used in recent field-theory inspired models \cite{2023PhRvL.130i1403I,2024PhRvD.109f4058S,2024PhRvD.110j3001S}. This problem seems eminently solvable, but as it was not the focus of the present effort we have not addressed it here.

\bibliography{TidalMatch}

\end{document}